\documentclass[aps,prb,reprint,superscriptaddress,showpacs, showkeys, showpacs]{revtex4-1}
\usepackage{hyperref}
\usepackage{graphicx}
\usepackage{amsmath}
\usepackage{color}
\usepackage{units}
\usepackage{tikz}
\usepackage{pgfplots}
\begin{document}

\title{Phase space manipulation of free-electron pulses from metal nanotips using combined THz near-fields and external biasing}%

\author{Lara Wimmer}
\affiliation{IV. Physical Institute - Solids and Nanostructures, University of G\"ottingen, Friedrich-Hund-Platz 1, 37077 G\"ottingen, Germany. }
\author{Oliver Karnbach}
\affiliation{IV. Physical Institute - Solids and Nanostructures, University of G\"ottingen, Friedrich-Hund-Platz 1, 37077 G\"ottingen, Germany. }
\author{Georg Herink}
\affiliation{IV. Physical Institute - Solids and Nanostructures, University of G\"ottingen, Friedrich-Hund-Platz 1, 37077 G\"ottingen, Germany. }
\author{Claus Ropers}
\email[]{claus.ropers@uni-goettingen.de}
\affiliation{IV. Physical Institute - Solids and Nanostructures, University of G\"ottingen, Friedrich-Hund-Platz 1, 37077 G\"ottingen, Germany. }

\date{December 15, 2016}%

\begin{abstract}
We present a comprehensive experimental and numerical study of photoelectron streaking at metallic nanotips using single-cycle Terahertz (THz) transients and a static bias voltage as an external control parameter. Analyzing bias voltage dependent streaking spectrograms, we explore the THz-induced reshaping of photoelectron energy spectra, governed by the superimposed static field. Numerical simulations are employed to determine the local field strengths and spatial decay lengths of the field contributions, demonstrating electron trajectory control and the manipulation of the phase space distributions in confined fields with both dynamic and static components.
\end{abstract}





\maketitle
\section{Introduction}
Progress in time-resolved experimental techniques is driving an increased understanding of the ultrafast electronic, structural and spin response of bulk materials, surfaces, and nanostructures. Ultrafast electron diffraction and microscopy \cite{Bos93, Siw03, Bau1, Gul14, LaG06, Fla10, Pia13, Fei16} provide unique insights into the dynamics of nanoscale processes, with broad applications ranging from the study of structural phase transitions \cite{Siw03, Eic10} to free-electron quantum optics \cite {Fei15, Ech16}. The capabilities of these techniques strongly depend on the quality of the electron source, including its spatial beam properties and pulse duration. While the transverse beam coherence is greatly enhanced by the use of nanoscale electron sources or tailored photoemission processes \cite{Hom061, Rop07, Bar09, Bor10,  Yan11, Paa12, Par12, Car14, Gul14, Krueger2014, Mul14, Bor15, Ehb15, Kea16}, ultrashort pulse durations are achieved either by minimizing propagation distances from the source to the sample \cite{Siw03, Gul14}, or by active means using time-dependent electric fields \cite{van07, van10, Gli12, Gli15, Cur16, Kea16}. Electron pulse compression in microwave cavities is an established technique to produce ultrashort pulses of high bunch charge \cite{van07, van10}.\\
At optical frequencies, inelastic near-field interactions with free electron beams \cite{Gar10, Bar09, Par10} were recently applied in a quantum-coherent manner \cite{Fei15, Ech16}, creating the possibility of forming attosecond electron pulse trains \cite{Fei15, Bau}. Between the optical and microwave domains, intense phase-stable mid-infrared and Terahertz waveforms currently promote access to nonlinear and strong-field phenomena \cite{Kampfrath2013, Leitenstorfer2014}, including Terahertz high-harmonic generation in semiconductors \cite{Schubert2014} and ultrafast scanning tunneling microscopy \cite{Cocker2013, Eisele2014, Cocker2016}. Moreover, mid-IR photoemission studies at metal nanotips \cite{Her, Par12, Car14} demonstrate the prospects for the manipulation of free-electron beams, yielding field-driven acceleration with unique sub-cycle dyanmics arising from the nanoscale field confinement. In strong-field photoemission from nanostructures \cite{Bor10,  Yalunin2011, Kruger2011, Zherebtsov2011, Krueger2012, Her, Yalunin2013, Dombi2013, Suessmann2015}, the emission and acceleration processes are driven by the same field and are thus closely coupled. Inducing both processes by independent fields, streaking spectroscopy allows for a characterization of electromagnetic fields by their impact on photoelectron spectra \cite{ Goulielmakis2004, Fruehling2009}. Originally developed in attosecond science and applied to gas phase targets and planar surfaces, the concept was recently transferred to optical nanostructures \cite{Sto07, Sussmann2011, Wim14}, demonstrating unique features offered by streaking in spatially-confined fields. In particular, THz near-field control and sub-cycle streaking of photoelectron emission from metal nanotips were realized \cite{Wim14}, followed by demonstrations of THz-induced tunneling emission \cite{Her14, Iwa15, Li2016}, electron acceleration in waveguides \cite{Hua15, Nan15, Fal16, Huang2016}, and electron pulse compression and deflection in resonant structures \cite{Kas12, Kel12, Fab14, Kea16, Rya16}. In the sub-cycle regime of electron acceleration, the energy gain in near-field streaking sensitively depends on the time spent by the electrons within the confined field \cite{Sto07, Her, Kel12, Wim14, Wimmer2016, Forg2016}. This suggests that an additional static field driving the electrons out of the localized streaking region may afford additional control over the electron dynamics.\\

Here, we experimentally demonstrate that an external bias voltage applied to a nanoscale photoemitter offers a powerful means of manipulating the electron dynamics in THz near-fields. The basic experimental concept is illustrated in Fig. 1. We conduct THz streaking spectroscopy of femtosecond photoelectron pulses emitted from the apex of a gold nanotip in the presence of a variable bias field, as depicted in Fig. 1a. The effect of the static bias on the electron dynamics illustrated in Figs. 1b-e. Generally, the negative bias field draws the electrons towards the detector (cf. Fig. 1b). For certain emission phases within the THz transient (cf. Fig. 1c), this additional acceleration strongly influences the electron trajectories and the photocurrent. For example, in emission phases of otherwise suppressed photocurrent, the negative bias field allows the photoelectrons to leave the THz near-field (see Fig. 1d). The bias-induced changes in the streaking spectrograms are illustrated in Fig. 2, showing drastic modifications of the time-dependent photoelectron spectra. Numerical simulations corroborating the experimental data allow for a quantitative characterization of the local field parameters of the THz and static fields. The analysis of ensembles of electron trajectories reveals that the phase space density distribution of the electron pulses is strongly affected by the combined action of the THz near-field and the bias.\\
Our work demonstrates the virtue of spatially confined electric fields in the THz range for the active control of ultrashort electron pulses with additional leverage obtained by a superimposed static field.

\begin{figure}
\includegraphics[width=\columnwidth]{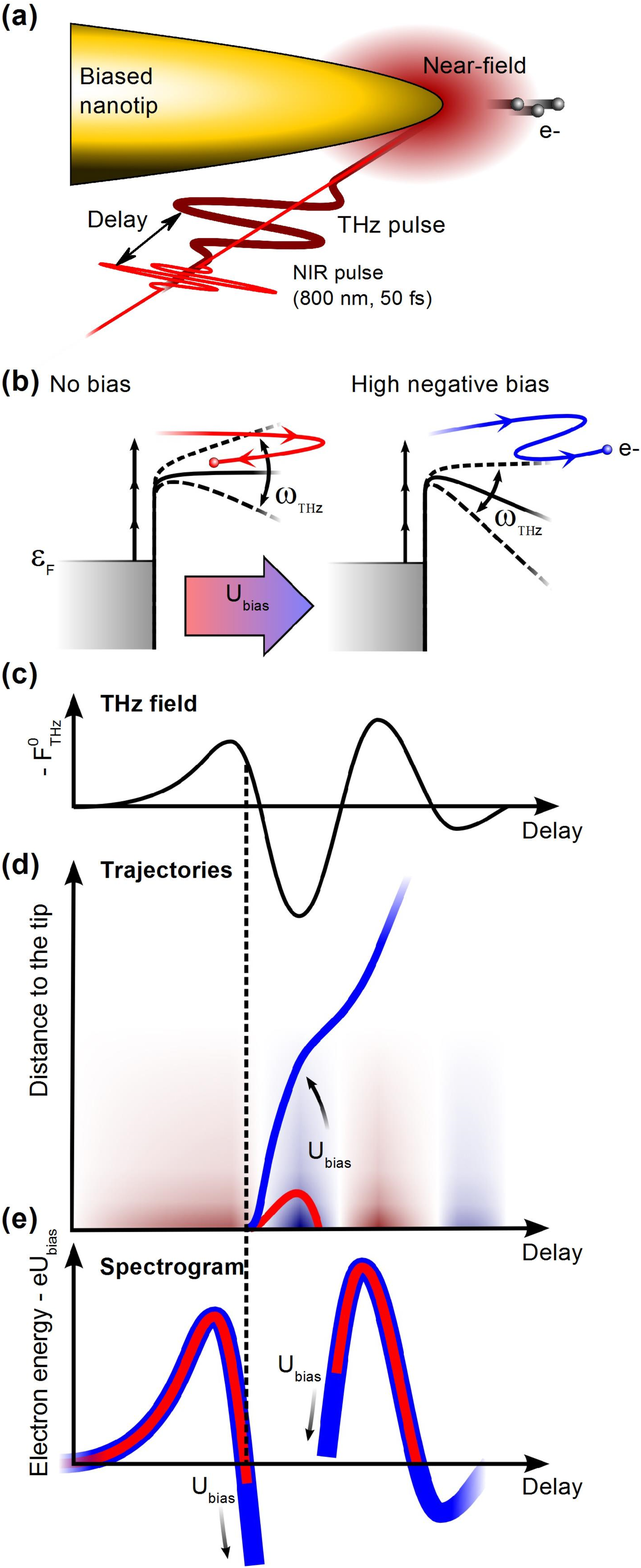}
\label{fig: 1}
\caption{Basic principle of THz streaking at biased nanotips. (a) Experimental streaking scheme. (b) Emission and propagation of photoelectrons in the absence of a bias (left, red trajectory) and with a high negative static bias (right, blue trajectory). The static field bends the potential, allowing for the electrons to leave the near-field. (c) THz surface electric field. (d) Electron trajectories without external bias (red trajectory) and with additional acceleration by a static bias field (blue trajectory). The background color indicates the THz electric field. (e) The sketched red and blue streaking spectrograms illustrate the impact of an increasing bias voltage.}
\end{figure}

\section{Experimental procedure and results}
In the experiments, 50-femtosecond near-infrared (NIR) pulses and single-cycle THz transients are jointly focused onto a gold nanotip (Fig. 1a). The NIR-pulse generates photoelectrons at the tip apex by multiphoton photoemission, which are subsequently accelerated in the THz-induced near-field. A negative bias voltage $U_{bias}$ applied to the tip additionally accelerates the electrons (cf. Fig. 1b). Kinetic energy distributions of photoelectrons are recorded with a time-of-flight spectrometer as a function of relative time-delay between the NIR and the THz pulses, resulting in streaking spectrograms as displayed, e.g., in Figs. 2a,b,e.\\
The streaking spectrograms demonstrate a strong influence of the THz electric field on the photoelectron emission and propagation. Before we discuss the influence of the external bias, let us first reconsider the main features of such spectrograms (see, e.g., Fig. 2a), some of which were covered in detail in our previous work \cite{Her, Wim14, Her14}.  The THz field affects both the photoemission current and the photoelectron kinetic energy. Generally, the tip exhibits a rectifying behavior, in which significant photocurrent is only observed for THz phases with positive force on the electrons, i.e., negative electric field, and the photocurrent is suppressed for an electric force directing the electrons to the tip (see, e.g., the current-free interval around zero delay in Fig. 2a). Upon variation of the delay, the kinetic energy spectra trace out the overall electron acceleration induced by the locally enhanced THz field and the static bias. Due to the high localization and field enhancement, which is characteristic for optically induced near-fields at metal nanostructures, the delay-dependent kinetic energy mainly follows the THz field in the moment of emission \cite{Wim14}.\\
The basic principles of field-driven electron acceleration in near-fields are characterized by a spatial adiabaticity parameter\cite{Her} $\delta=l_f/l_q$, which relates the spatial near-field decay length $l_f$ with the electron quiver amplitude $l_q=eF/m_e\omega^2$ in a (homogeneous) field of strength $F$ and at a frequency $\omega$. Field-driven, sub-cycle dynamics, in which photoelectrons leave the optical near-field directly after the emission process, correspond to $\delta\ll1$, whereas electron dynamics governed by propagation effects are associated with a larger $\delta$. In the present case of studying the electron dynamics in a phase-resolved manner, we can identify the maxima of the kinetic energy trace with field-driven electron dynamics. While the streaking waveform is generally characterized by quasi-instantaneous electron acceleration \cite{Wim14}, in emission phases near zero-crossings of the THz field, propagation effects are expected to have a measurable influence. Around these phases with near-zero streaking field, the electron kinetic energy sensitively depends on the propagation time within the THz near-field, which can be controlled by the external bias.\\
In order to investigate the effect of a static field on the streaking process, we systematically record a series of streaking spectrograms for bias voltages $U_{bias}$ in the range of -20 to -150 V, for fixed optical and THz excitations.\\
The NIR beam ($40\:\mu$W input power, $10$ mm beam diameter) is focused by a $150$ mm planoconvex lens and the THz beam ($20$ mm diameter) is focused by a parabolic mirror (focal length of $25.4$ mm) to about $1$ mm. All measurements presented in this work are recorded using the same nanotip with an apex radius of curvature of $20$ nm (cf. Fig. 6). The main trends observed in this series (Fig. 2e) are evident by the comparison of two exemplary spectrograms at -20 V and $-100$ V bias (Figs. 2a and 2b, respectively). We focus our discussion on three characteristic features, which are the maximum energy $E_{max}$ at $t_0=0.4$ ps, the minimum energy $E_{min}$ near the emission phase of $t_0\approx0.2$ ps, and the energy level $E_{on}$ at the "onset" of the photocurrent at $t_0\approx0.2$ ps after the interval of suppressed photocurrent, as marked in Fig. 2d. We will use these features to quantitatively determine the strength and nanoscopic spatial profile of the electric field surrounding the tip.\\ 
The most prominent differences between the streaking spectrograms at $-20$ V and $-100$ V bias appear at $E_{min}$ and $E_{on}$. With increasing negative bias voltage, both energies decrease relative to the contribution of the bias voltage, $-eU_{bias}$ (dashed line in Figs. 2a-d). Whereas the minimum kinetic energy essentially coincides with the bias energy at $U_{bias}=-20$ V (cf. Fig. 2a), it drops by more than $10$ eV below the bias energy level for $U_{bias}=-100$ V (cf. Fig. 2b).\\
The bias-dependencies of $E_{max}$, $E_{min}$, and $E_{on}$ are displayed in Fig. 2f, extracted from the series of measurements in Fig. 2e, and offset by the bias energy. The maximum THz-induced energy gain ($E_{max} +eU_{bias}$) is nearly constant at about $26$ eV for all bias voltages (black squares). In contrast, both the minimum and the onset energies (blue and red squares, respectively) depend linearly on bias voltage with a common slope of $0.135$ eV/V$\cdot U_{bias}$, such that $E_{min}+eU_{bias}=0.135\: \unit{eV/V}\cdot U_{bias}$ and $E_{on}+eU_{bias}=19 \unit{eV}+0.135 \unit{eV/V}\cdot U_{bias}$.

\begin{figure*}
\includegraphics[width=1\linewidth]{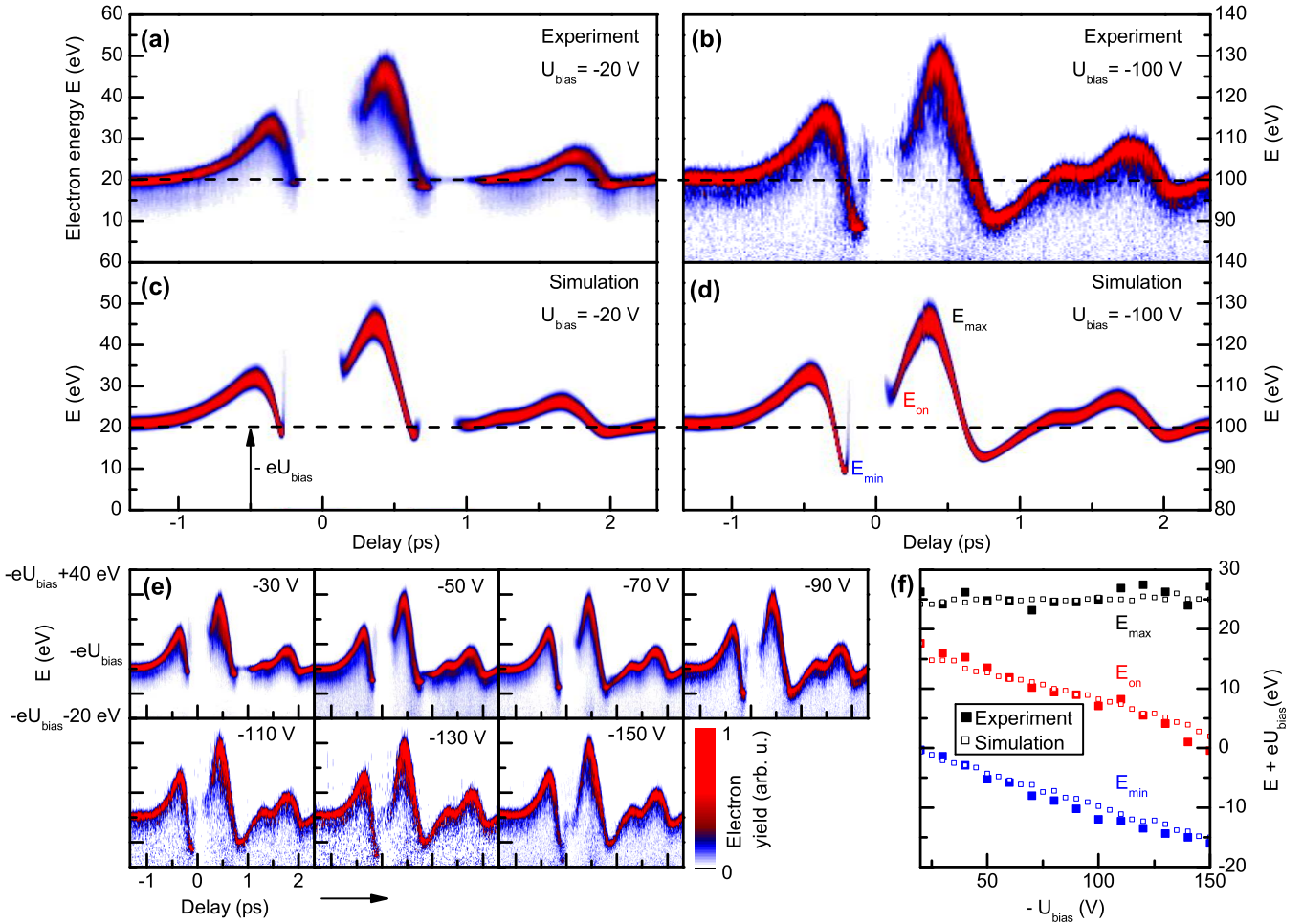}
\label{fig: 2}
\caption{THz streaking spectrograms at biased metal nanotips. (a), (b) Experimental streaking spectrograms (bias voltage:  $- 20$ (a) and $-100$ V (b)). (c) and (d): Corresponding simulations of the streaking spectrograms. (e): Extended series of streaking spectrograms recorded at bias voltages ranging from - 30 V to - 150 V. (f) Extracted bias dependence of the minimum and maximum energy gain in the THz-field, and of the energy gain at the onset of the photocurrent (see labels in (d)).}
\end{figure*}

\section{Numerical simulation of streaking spectrograms}
Obtaining a quantitative understanding of nanostructure streaking in the presence of static fields requires a detailed consideration of the joint action of the superimposed static and dynamic fields on the photoelectrons. The final kinetic energy $E_{end}$ of a photoelectron with initial kinetic energy $E_0$ is given by the integral over the electric field along the trajectories $s(t)$ corresponding to the time-dependent distance from the tip surface:

\begin{equation}
E_{end}=E_0-e\int_0^{\infty}\left[F_{stat}(s)+F_{THz}\left(s(t), t\right)\right]ds.
\label{eqn: 1}
\end{equation}

The electric field at the nanotip apex is the superposition of the static field $F_{stat} (s)$  and the time-dependent THz near-field $F_{THz} (s(t),t)$. The THz-induced energy gain, i.e., the second term in the integral, depends on the individual trajectories $s(t)$. Besides the initial kinetic energy $E_0$ and the emission time $t_0$, these trajectories depend on the bias voltage $U_{bias}$. In particular, the strength of the static field driving the electrons away from the tip determines their effective interaction time with the THz near-field. Predicting the quantitative influence of the bias field on the streaking process, however, warrants numerical modeling.\\
In our simulations of streaking spectrograms, we describe the THz field by its field strength $F_{THz}^0$ at the tip surface ($s=0$) and a dipolar spatial decay length $l_f$ (see Appendix \ref{par: B}). The local THz waveform is determined from experimental data (see Appendix \ref{par: B}). The static electric field is modeled as a superposition of a homogeneous long-range component with field strength $F_{stat}^1$ and a more confined, tip-induced component with a surface field of $F_{stat}^0$ and decay length $l_f$ also used for the THz near-field. The variables for the static field are constrained to satisfy the total energy $-eU_{bias}$ gained by the bias voltage (first term in the integral in Eq. \ref{eqn: 1}).\\
In analogy to the widely-used simpleman's's's's model of strong-field photoemission from atoms \cite{Cor93, Pau95} or metallic nanotips \cite{Krueger2012, Her}, the simulation of the spectrograms is composed of two steps, namely (i) the photoemission process and (ii) the field-driven propagation of photoelectrons. The photoemission is described as a multiphoton process using the Fowler-Dubridge model \cite{Fow31, DuB331} and includes the reduction of the effective work function via the Schottky effect \cite{Sch23}. For each delay, the field-dependent initial electron energy spectrum at the surface is taken into account prior to the propagation in the THz near-field. The propagation step computes point-particle electron trajectories in the spatio-temporally varying field composed of the static and THz-frequency components (see Appendix \ref{par: B}).\\
From the simulations, we generally find that the delay-dependent photocurrent modulation and the energy distribution are governed by both the field-dependence of the emission probability and propagation effects. Whereas the enhancement of the photocurrent is an instantaneous effect due to Schottky barrier-reduction by the static and THz fields, the suppression of the photocurrent results from the propagation of photoelectrons back to the tip.

\begin{figure}
\includegraphics[width=\columnwidth]{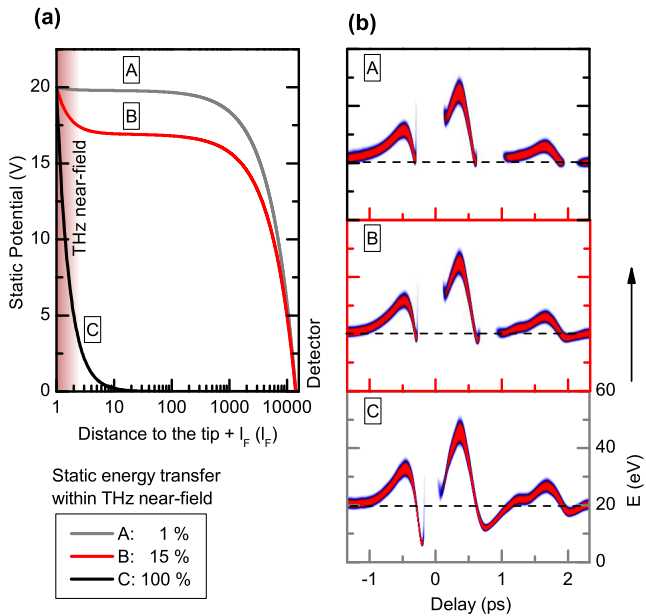}
\label{fig: 3}
\caption{Influence of the static bias onto the streaking spectrograms. (a) Spatial decay of the static potential for three scenarios. The red-shaded area represents the THz near-field. A: 1 \%-fraction of $-eU_{bias}$ is gained on the scale of the THz near-field. B: 15 $\%$-fraction C: At 100 $\%$, the electron is accelerated to its final velocity within the THz near-field. (b) Simulated streaking spectrograms for cases A-C.}
\end{figure}

\section{Influence of the near-field parameters on the electron dynamics}
The streaking measurements contain direct information on the spatial decay of the static and THz fields close to the nanotip apex (see Appendix \ref{par: B}). For example, the energies $E_{min}$ and $E_{on}$ measured in the streaking spectrograms reflect the degree, to which the static potential drops within the confined region of the THz near-field. Figure 3 illustrates the change in the computed streaking spectrograms for three scenarios in the spatial profile of the bias field. These cases correspond to an essentially homogeneous static field (A, no lightning-rod field enhancement), a fully localized static field with a sharp drop over the distance $l_f$ (C), or a mixed scenario (B).\\
If the electrons acquire the bias energy essentially outside the near-field region (A), the streaking waveform is merely shifted by the bias and is restricted to the accelerating half-cycles of the THz transient, and $E_{min}\approx -eU_{bias}$. The situation becomes more complex if there is a considerable decay of the bias field within the THz near-field region, illustrated in (B, C). Here, the minimum energy is influenced by the temporal evolution of the THz transient, and one finds kinetic energies below the bias energy, i.e., $E_{min}<-eU_{bias}$.\\
Let us now relate these simulated trends to the experimental observations. At low bias voltages (cf. Fig. 2a, -20 V), the minimum kinetic energy in the streaking spectrograms largely coincides with the bias energy level. With increasing acceleration in the negative bias field, the minimum energy decreases relative to the bias energy level by about 0.135 eV per Volt applied negative bias (cf. Fig. 2f). To account for this scaling, we simulated streaking traces in a static potential partially decaying within the THz near-field (cf. Fig. 3(B) and Figs. 2c,d). The series of measurements displayed in this work is best reproduced by a minimum surface THz field strength of $F_{THz}^0=-3.15$ MV/cm and a THz decay length of $l_f=55$ nm. Considering the incident THz field strengths of $40$ kV/cm, we estimate a field enhancement factor about 80. The static energy gain within the THz near-field amounts to $15.7$ eV for $- 100$ V bias voltage, corresponding to a static surface field of $F_{stat}^0=-1.45$ MV/cm. This energy gain is relatively independent of the specific functional form of the bias field.\\
Figures 2c,d show simulated streaking spectrograms for a low ($- 20$ V) and a high ($-100$ V) negative bias voltage, respectively, for direct comparison to the experimental measurements in Figs. 2a,b. We simulate the experimental streaking spectrograms with one universal set of parameters for all bias voltages. Overall, these simulations are in remarkable agreement with the measured spectrograms. Specifically, our simulations reproduce the experimentally observed scaling of the characteristic energies $E_{max}$, $E_{on}$, and $E_{min}$, as shown in Fig. 2f. The energy gain in the THz field at $E_{max}$ is independent of bias voltage, which is a clear sign of field-driven dynamics. The linear decrease of the energies $E_{on}$ and $E_{min}$ result from the decay of the static potential within the THz near-field (specifically, 15 \% of the static potential decays within $l_F$). The full set of spectrograms is provided in the Appendix.

\begin{figure*}
\includegraphics[width=1\linewidth]{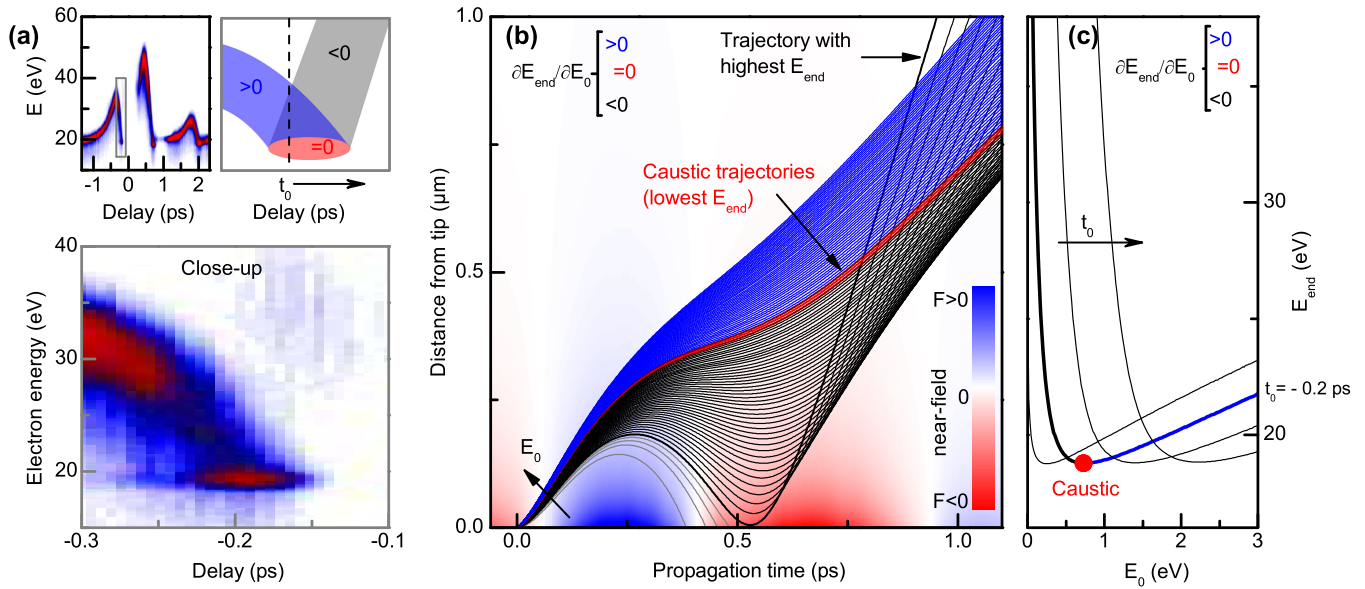}
\label{fig: 4}
\caption{Caustic trajectories at $-20$ V bias voltage. (a) Experimental streaking trace with a close-up at emission times directly before a complete suppression of the photocurrent. (b) Simulated trajectories with different initial energies. The background colors show the time-dependent near-field. The colors of the trajectories refer to the respective sections in (c). (c) Final energy as a function of initial energy (transfer function) featuring a minimum in the final energy (caustic, red). Emission for time $t_0$ and additional curves for delays in  steps of $10$ fs.}
\end{figure*}

\section{Caustic trajectories}
The numerical simulations reproduce fine details of the experimental spectrograms, and allow for an in-depth analysis of  the underlying computed electron dynamics and trajectories. For example, we will discuss the minimum energy feature prior to the gap around $t_0=-0.2$ ps (see Fig. 4a). The characteristics of this spectral feature are a high electron yield at the energy minimum and a smaller number of electrons with higher energy. The minimum energy is rather well-defined and nearly constant over a delay interval of around $100$ fs.\\
This spectral feature implies a considerable THz-induced spectral reshaping of the initial photoelectron spectrum \cite{Wim14}. We investigate this characteristic detail by simulating electron trajectories for a distribution of initial kinetic energies, as shown in Fig. 4b. The final kinetic energy of the photoelectrons as a function of their initial energy is presented in Fig. 4c for different emission times. This plot depicts the energy transfer function relating the final energy $E_{end}$ to the initial energy $E_0$. The slope of $E_{end}(E_0)$ characterizes three different classes of trajectories, which are marked by the color-code in Fig. 4a. In case of high initial energies (marked in blue), $\partial E_{end}/\partial E_0$ is positive and asymptotically approaches unity, indicating quasi-static electron acceleration with the THz-induced energy gain being independent of the initial energy. As illustrated in Figure 4b, the electrons with the lowest initial energy are too slow to escape the positive decelerating THz near-field (grey trajectories), which leads to a reduction of the total photocurrent. A minimum initial energy is required to escape the near-field. This critical initial energy corresponds to a grazing trajectory and the maximum final energy (thick black line). Upon further increasing the initial energy, the reacceleration in the negative half-cycle occurs at larger distances from the tip. Hence, higher initial energies lead to lower final energies, and these trajectories (black lines) exhibit a negative slope of the transfer function, $\partial E_{end}/\partial E_0<0$.\\
At the transition between both classes of trajectories (blue and black), the transfer function changes sign (red) and its slope approaches zero. In particular, this minimum of the transfer function $E_{end} (E_0)$ leads to the observed caustic feature in the spectrogram, which is defined by the mapping of different trajectories onto the same final energy ($\partial E_{end}/\partial E_0=0$). Notably, the respective initial energies of the caustic energy minimum shift with the delay from $E_0=0$ to higher initial energies, as shown in Fig. 4c. Thus, the range of delays with caustic behavior is closely related to the width of the initial spectrum.\\
More generally, caustic trajectories occur for an appropriate combination of a rapid spatial field decay and a fast temporal rise in the accelerating field. Such a situation is also found at the onset at the second accelerating half-cycle at $t_0=0.2$ ps; see Appendix, Fig. 9.

\begin{figure*}
\includegraphics[width=1\linewidth]{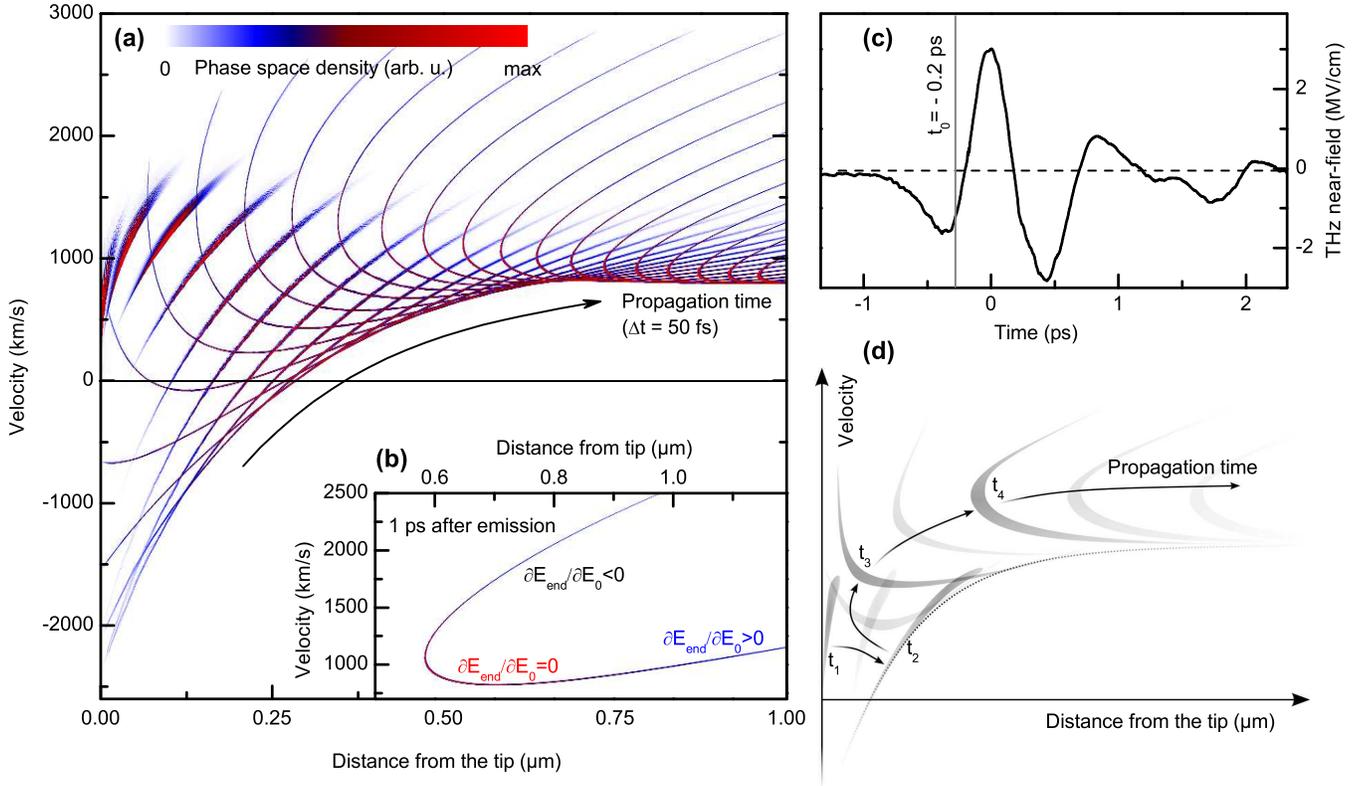}
\label{fig: 5}
\caption{Simulated caustic trajectories in phase space (same emission time $t_0= - 0.2$ ps and bias voltage $- 20$ V as in Fig. 3). (a) Phase space density distribution for different propagation times after photoemission (step width $\Delta t=50$ fs, weighted with the incident spectral distribution and averaged over emission times with an FWHM of $30$ fs). (b) Exemplary phase space density distribution $1$ ps after emission. The caustic trajectories show the lowest kinetic energy. (c) The near-field at the tip surface with the emission time marked in grey. (d) Sketch illustrating the temporal evolution of the phase space density.}
\end{figure*}

\section{Evolution in phase space}
Trajectories and transfer functions (cf. Fig. 4) provide a mapping of initial to final kinetic energies. However, to numerically evaluate the spectral reshaping and the longitudinal electron pulse properties, we incorporate the initial energy spectra and analyze the resulting time-dependent electron distribution in phase space.\\
The temporal evolution of the caustic trajectories at $t_0=- 0.2$ ps (cf. Fig. 4) is depicted in phase space in Fig. 5. Figure 5a displays snapshots ($\Delta t=50$ fs) of the phase space density. Each snapshot is a scatter plot of space and velocity coordinates, extracted from the trajectories at a specific propagation time. The individual data points are weighted with the electron yield for the initial energy of the respective trajectory (further details given in Appendix \ref{par: C}). The initial electron distribution is first stretched and then warped close to the tip surface. After leaving the THz near-field, the underlying convex transfer function is reflected in the curved phase space density distribution (cf. Fig. 4c) as depicted in Fig. 5b: The lower-velocity branch corresponds to the initially faster electrons with $\partial E_{end}/\partial E_0>0$  (blue trajectories in Fig. 4b), and the higher-velocity branch is associated with the black trajectories in Fig. 4b. In Figure 5b, the caustic electrons exhibit the lowest final momentum.\\
Generally, the manipulation of the phase space density distribution sensitively depends on the emission time $t_0$ into the THz field. Electrons emitted during increasing acceleration, e.g., at a delay of $t_0=0.2$ ps, experience a spectral compression, as well as an overall contraction in phase space in comparison to static acceleration (cf. Fig. 9, Appendix \ref{par: C}).\\
The modification of the populated longitudinal phase space volume is enabled by the spatio-temporal variation of the THz near-field leading to a trajectory-dependent energy gain $E_{THz}$. This contrasts to conservative forces, which necessarily preserve the populated phase space volume (Liouville's theorem).

\section{Conclusions}
In summary, we found that an external bias voltage is a powerful control parameter in streaking spectroscopy from contacted metallic nanostructures, which impacts ultrafast electron dynamics at the nanoscale. Both the waveform and the spatial profile of THz streaking fields can be extracted from comparing numerical simulations with characteristic features in experimental spectrograms. Caustic trajectories, which are associated with nearly grazing return trajectories at the surface, are found to be particularly sensitive to the THz and static field distribution. The spatio-temporal electron dynamics in THz near-fields allows for a tuning of electron energy and phase space distributions. Our work highlights emerging fundamental prospects for ultrafast electron pulse control arising from the combined action of static and time-dependent near-fields.

\begin{table}[h]
\begin{tabular}{ll}
\hline\hline
$E_0$ & Initial electron energy after multiphoton emission\\
$E_{end}$ & Final electron energy at the detector\\
$E_{max}$ & Maximum electron energy\\
$E_{min}$ & Minimum electron energy($t_0\approx-0.2$ ps)\\
$E_{on}$ & Electron energy at the onset ($t_0\approx0.2$ ps)\\
$F_{stat}^0$ & Static surface electric field (dipole component)\\
$F_{stat}^1$ & Static homogeneous electric field component\\
$F_{THz}^0$ & Amplitude of the THz surface electric field\\
$U_{bias}$ & Static bias voltage\\
$e$ & Elementary charge\\
$\delta$ & Spatial adiabaticity parameter\\
$l_f$ & Field decay length\\
$l_q$ & Electron quiver amplitude\\
$m_e$ & Electron mass\\
$\omega$ & THz frequency\\
$s$ & Distance to the tip\\
$t_0$ & Emission time\\
\hline\hline
\end{tabular}
\caption{\label{tab: 1}Notations and abbreviations.}
\end{table}

\appendix
\section{Experimental methods and data analysis}
\label{par: A}
An extended series of experimental streaking spectrograms is presented for additional bias voltages in Fig. 6. Each spectrogram is normalized to the maximum electron spectral density at large negative delay. For comparison, all streaking spectrograms use a common color scale.\\
The kinetic energy distributions of the photoelectrons are detected with a time-of-flight electron spectrometer. The energy resolution is given by the temporal resolution of the acquisition electronics, and varies with the kinetic energy of the photoelectrons. In order to maintain constant energy resolution for different bias potentials, we apply a negative drift voltage of $U_{drift}=U_{bias}+30$ V.

\begin{figure*}
\includegraphics[width=0.925\linewidth]{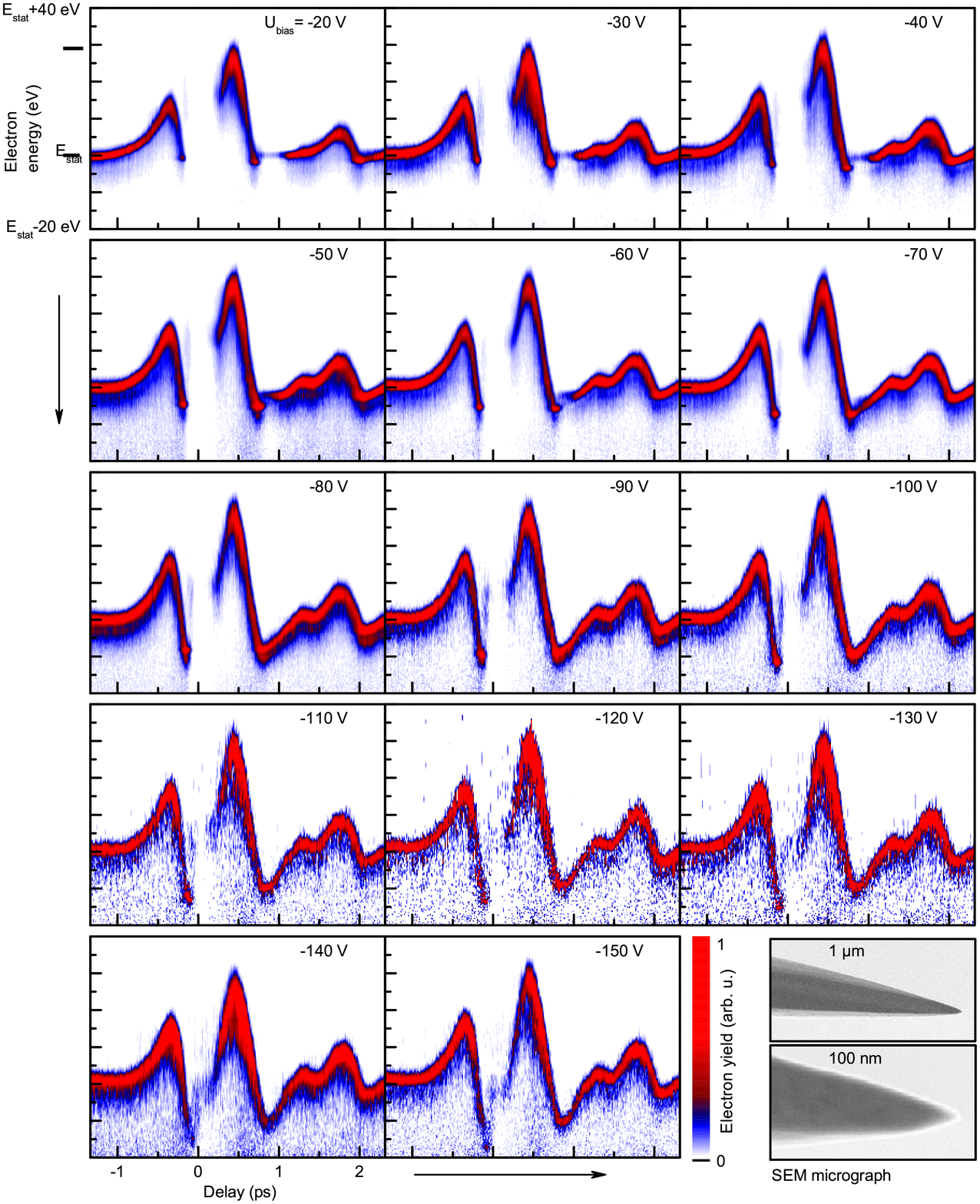}
\label{fig: 6}
\caption{Complete series of experimental streaking spectrograms for bias voltages from $- 20$ to $- 150$ V and SEM micrographs of the gold nanotip used to record the streaking series.}
\end{figure*}

\section{Numerical simulations and analytical streaking model}
\label{par: B}
In this section, we describe the numerical simulation of the spectrograms in detail, and we introduce an analytical description of the electron dynamics observed in our experiments. Both the numerical and the analytical model are based on the same spatial THz near-field profile presented in the following. For the numerical simulations, we focus on the two-step-model of photoemission and propagation of electrons and elucidate our procedure to extract the near-field parameters from experimental data.

\begin{figure*}
\includegraphics[width=1\linewidth]{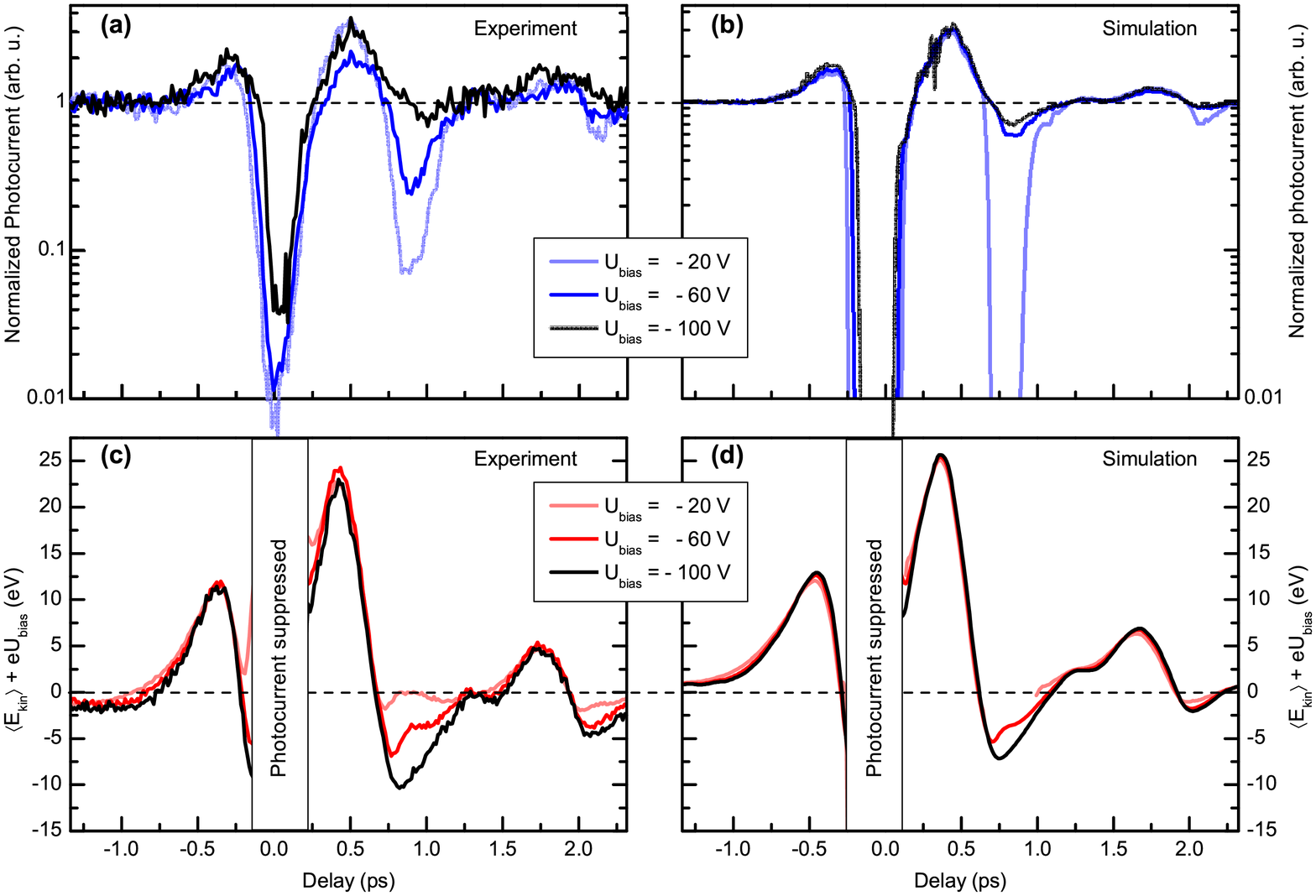}
\label{fig: 7}
\caption{Influence of a static bias voltage onto photocurrent and kinetic energy. (a) and (b) Photocurrent, measured (a) and simulated (b) for bias voltages of $- 20$ V, $- 60$ V, and $- 100$ V.  (c) and (d): Measured and simulated energy expectation values for  same bias voltages. During the intervals of suppressed photocurrent, the assignment of kinetic energies is not possible.}
\end{figure*}

\subsection{Spatial and temporal dependence of the near-field}
As discussed in the main text, the electric field surrounding the nanotip apex is modeled consisting of two components: a static field induced by the negative bias voltage and the THz-induced near-field, which varies in time. Both components decay on the scale of the apex curvature. \\
We describe the spatial dependence of the THz-induced near-field $F_{THz} (s,t)$ by a dipolar field, characterized by a decay length $l_f$, defined by $F_{THz} (l_f )=F_{THz}/2$. For convenience, we introduce a rescaled parameter $d$,  $l_f=(\sqrt[3]{2}-1)\cdot d$:

\begin{equation}
F_{THz}(s,t)=F_{THz}^0(t)d^3/\left(d+s\right)^3.
\label{eqn: 2}
\end{equation}

To construct the local THz near-field field $F_{THz}^0 (t)$, we employ a combination of the energy expectation value at a bias voltage of $-100$ V and the incident electric field measured by electro-optic sampling (EOS). The expectation value at high negative bias represents a reliable measure of the momentary field at all times outside the suppression of the photocurrent in one half-cycle. Figure 8e compares the energy expectation value with the incident electric field, demonstrating some ringing of the near-field after the initial cycle (compare also Ref. \cite{Wim14}). However, this frequency-dependent response has a weaker impact on the main cycle and the half-cycle not captured by the energy expectation value. We use this observation to complete the near-field in the missing half-cycle by replacing it with a scaled half-cycle determined by EOS.\\
A small delay difference of $80$ fs between the maximum of the photocurrent (corresponding to the maximum of the near-field strength) and the maximum energy of the streaking waveform is caused by the propagation in the THz near-field. This time difference is well-reproduced in our numerical simulations. \\
In our model, the static field is composed of the two spatial components, such that the total static field $F_{stat}$ (s) is given by:

\begin{equation}
F_{stat}(s,t)=F_{stat}^0d^3/\left(d+s\right)^3.
\label{eqn: 3}
\end{equation}

As introduced in the main text, $F_{stat}^1$ is the long-range component of the bias field, and $F_{stat}^0$ is the electric field strength of the dipolar component at the surface.

\subsection{Photoemission process}
The NIR-induced photoemission is described as a multi-photon process following the Fowler-DuBridge theory \cite{Fow31, DuB331, DuB33, Bechtel1977, Brogle1996}. It is based on the Sommerfeld model of a free electron gas and one-dimensional emission. The energy difference between the sum of n-photon energies and the initial state in the metal determines the excess kinetic energy of the emitted electrons. The relative contributions of the different emission channels are determined using the spectrally integrated photocurrent trace at $U_{bias}= - 100$ V. To include the Schottky effect, we employ an effective work function \cite{Sch23} $\Phi_{eff}=\Phi-\sqrt{\frac{e^2F}{4\pi\varepsilon_0}}$. The model qualitatively reproduces the THz-induced photocurrent modulations for all bias voltages (cf. Fig. 7). Considering the overall spectral resolution in the experiments, we do not include a more detailed description of the photoemission process or carrier relaxation at the surface.

\subsection{Propagation of photoelectrons}
The electron trajectories are computed in a spatially and temporally varying one-dimensional electric field using a Runge-Kutta scheme for various delays between NIR and THz pulses, and are weighted by the respective initial energy spectra. This yields the final electron spectra of the streaking spectrograms. \\
For better comparison between measured and simulated streaking spectrograms, we convolve the numerical results with the energy resolution of the electron spectrometer.

\subsection{Extraction of the near-field parameters}
The near-field parameters are obtained via best agreement with the experimental streaking spectrograms. The maximum energy gain in the THz field $E_{max}+eU_{bias}$ is determined by $F_{THz}^0$, and $l_f$, and is independent of the bias voltage $U_{bias}$, which indicates field-driven acceleration of photoelectrons.\\
This set of optimal parameter values allows us to reproduce the photocurrent modulation, as well as the kinetic energy trace as shown in Fig. 7. The photocurrent (cf. Fig. 7a,b) is essentially in-phase with the streaking waveform (cf. Fig. 7c,d), as expected for field-driven electron dynamics \cite{Wim14}. The photocurrent, depicted in Fig. 7a, changes from a complete suppression to a strong enhancement at maximum THz fields. While the photocurrent enhancement is not particularly sensitive to the bias voltage in the range studied here, the influence of the static field is strongest close to the zero-crossings of the surface electric field due to a fraction of the photoelectrons propagating back to the metal surface. \\
We note that the experimental streaking spectrograms could also be reproduced by a somewhat different spatial dependence of the local contribution to the static field, e.g., by an exponential decay. Physically, the static energy gain of the electrons within the THz near-field governs the propagation effects and is thus the universal property identified.
However, the general parameters are obtained are very similar for different functional forms. As an illustration of the uniqueness of the parameters obtained, we fit the simulations with respect to the minimum and maximum kinetic energy $E_{min}$ and $E_{max}$ alone, which allows for different spatial decay lengths (cf. Fig. 8). For a low surface field strength and a long spatial field decay, the interaction time between the THz electric field and the photoelectrons increases, shifting the photocurrent onset to higher energies. Thus, only one set of parameters matches to the measurement in all three energies $E_{max}$, $E_{min}$, and $E_{on}$.

\begin{figure*}
\includegraphics[width=1\linewidth]{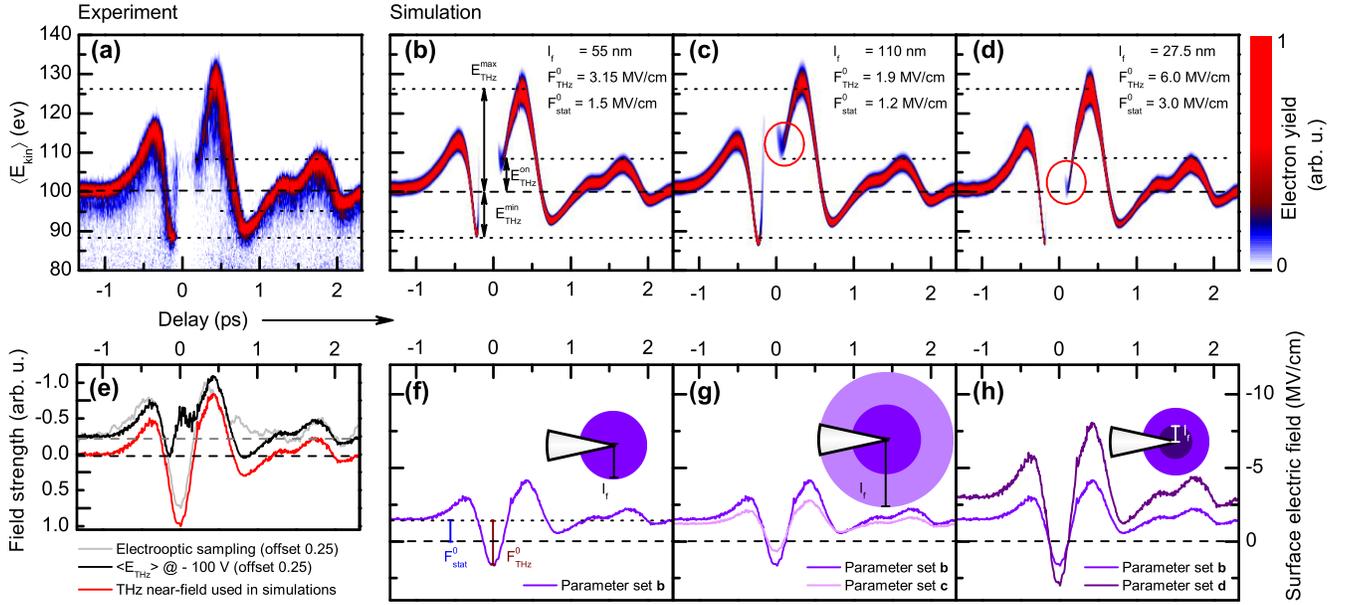}
\label{fig: 8}
\caption{Influence of the near-field parameters on the streaking spectrograms. (a) Measured spectrogram at $-100$ V bias. (b) Simulated spectrogram with optimized parameters. (c) Simulated spectrogram with larger decay length of the THz field ($l_f$ doubled). $F_{THz}^0$ and $F_{stat}^0$ are used to adjust maximum and minimum kinetic energies. The resulting "onset" energy $E_{on}$ of the photocurrent at $t_0=0.2$ ps (cf. red circle) is too high. (d) Simulated spectrogram with a very rapidly decaying THz field ($l_f$). Here, the onset energy $E_{on}$ is too low (cf. red circle).  (e) The incident electric field used in the simulations is extracted from the energy expectation value and electro-optic sampling. (f) Temporal evolution of the total surface electric field applied in simulation b. The sketch illustrates the spatial decay of the electric field. (g) and (h): Analogous illustration for the simulations shown in (c) and (d), respectively.}
\end{figure*}

\subsection{Analytical description}
The following analytical model facilitates an intuitive physical understanding of the electron dynamics. An analytical solution for the trajectories in the spatio-temporally varying field is not directly possible; we therefore introduce several approximations: The influence of the bias onto the streaking spectrograms is described by a set of analytical equations for the three energies $E_{max}$, $E_{min}$, and $E_{on}$. We use a spatial decay of the static and the THz electric field analogous to our numerical model. The temporal oscillation of the driving field is given by a $\cos(\omega t)$ term:

\begin{equation}
E_{end}=e\int_0^{det}F_{THz}^0(\omega t)\frac{d^3}{(d+s)^3}-F_{stat}^0\frac{d^3}{(d+s)^3}-F_{stat}^1.
\label{eqn: 4}
\end{equation}

As discussed earlier, electrons at the maximum energy of the streaking trace ($t_0=\pi/\omega$) exhibit quasi-static field-driven dynamics, and the time-dependence of the electric field can be neglected in computing the final energy. We obtain for the maximum energy gain in the THz field:

\begin{equation}
E_{max}+eU_{bias}=\frac{eF_{THz}^0d}{2}.
\label{eqn: 5}
\end{equation}

At the onset of the photocurrent at $t_0\approx\pi/2\omega$ ($t_0\approx0.2$ ps in the experiment), the time dependence of the electric field is approximated by a Taylor expansion of first order, and we obtain the THz-induced onset energy component:

\begin{equation}
E_{on}+eU_{bias}=-e\int_0^{\infty}\left(F_{THz}^0\omega\tau-F_{stat}^0\right)\frac{d^3}{(d+s)^3}ds.
\label{eqn: 6}
\end{equation}

Via the definition of the onset, the total surface electric field in the moment of emission is zero ($0=F_{stat}^0+F_{THz}^0 (t_0)$). As the respective energy is evaluated for the THz near-field, the contribution due to acceleration in the homogeneous static field $F_{stat}^1$ is neglected. The integral becomes:

\begin{equation}
E_{on}+eU_{bias}=-\frac{ed}{2}\left(F_{THz}^0\omega\tau-F_{stat}^0\right).
\label{eqn: 7}
\end{equation}

An increased static field shifts the emission time at the onset to phases of negative THz force (positive THz field strength). As the temporal slope of the THz transient at the surface is constant, the propagation time $\tau$ in the near-field and, thus, the THz-induced energy gain also remains constant ($\approx19$ eV). Thus, the external bias does not change the trajectories at the onset. The final kinetic energy of the photoelectrons results from this near-field energy component ($\approx19$ eV) and the the static long-range component. \\
The observed decrease of the THz energy gain $E_{on}+eU_{bias}$   ($0.135\:\unit{eV/V}\cdot U_{stat}$) at the onset arises from a decreased THz-acceleration and corresponds to the energy transferred by the rapidly decaying static field component. \\
The same linear dependence of this onset energy (at $t_0=0.2$ ps) is observed at the energy minimum for $t_0=-0.2$ ps (i.e. $t_0\approx\pi/2\omega$): The kinetic energy directly after the escape from the THz near-field amounts to nearly zero, leading to the following equation for the energy minimum:

\begin{equation}
E_{min}+eU_{bias}=\frac{eF_{stat}^0d}{2}
\label{eqn: 8}
\end{equation}

In this approximation, the final energy results from the spatial integral over the homogeneous component of the static field alone. This value thus represents the lower limit of the minimum electron energy. Remarkably, the deviation between the numerical simulation and the analytical approximation is less than $2$ eV. \\
At this point, the three equations \ref{eqn: 5}, \ref{eqn: 7}, and \ref{eqn: 8} describe three essentiall energies, which yield the near-field parameters from experimentally found energies:

\begin{itemize}
\item Maximum: \\ $E_{max}+eU_{bias}=\frac{eF_{THz}^0d}{2}=26$ eV
\item Onset: \\ $E_{on}+eU_{bias}=-\frac{ed}{2}\left(F_{THz}^0\omega\tau-F_{stat}^0\right)=19\:\unit{eV}+0.135\:\unit{eV/V}\cdot U_{bias}$
\item Minimum: \\  $E_{min}+eU_{bias}=\frac{eF_{stat}^0d}{2}=0.135\:\unit{eV/V}\cdot U_{bias}$
\item Static contribution: \\ $E_{stat}=-e\left(\frac{F_{stat}^0d}{2}+F_{stat}^1d_{det}\right)=-eU_{bias}$
\end{itemize}

$d_{det}=3$ mm is the distance from the tip to the detector. Solving this system of equations for a center frequency of $\omega=2\pi\cdot1.2$ THz leads to $F_{stat}^1=290$ V/cm, $F_{stat}^0/F_{THz}^0=0.51$ and $\tau=96$ fs. This value relates to the experimentally observed temporal shift of 80 fs between the maximum of the photocurrent and the kinetic energy trace.\\
Whereas the system of equations describes our experimentally observed bias-impact onto the streaking spectra, they still do not allow for an explicit solution of the near-field parameters. However, using the extracted parameters, we can determine an upper limit for the spatial decay length $l_f$ based on the acceleration to $19$ eV at the photocurrent onset: $d=\tau\sqrt{2\cdot19 \unit{eV}/m_e}\approx250$ nm, corresponding to a decay length of $l_f=65$ nm.

\begin{figure*}[htb]
\includegraphics[width=1\linewidth]{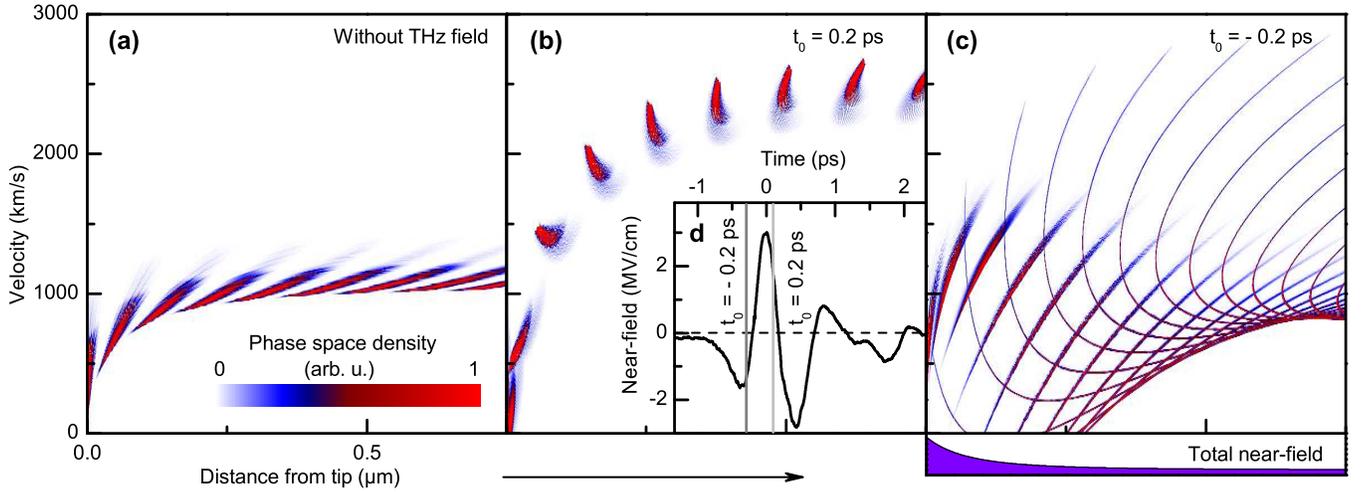}
\label{fig: 9}
\caption{Time evolution of the phase space density distribution simulated for different emission times at $-20$ V bias voltage. (a) Evolution of the phase space distribution of an electron pulse accelerated by the static field of the bias voltage alone ($\Delta t=100$ fs). (b) Phase space density for emission time $t_0=0.2$ ps. Snapshots taken with time differences of $\Delta t=50$ fs. Note the caustic flipping of the phase space distribution at velocities of $\approx 1300$ km/s. (c) Phase space density distribution for  $t_0=-0.2$ ps with $\Delta t=50$ fs as shown in Fig. 5 (negative velocities cropped in the diagram). The spatial decay of the near-field (absolute value of the maximum THz field, no time dependence) is depicted below (c). (d) Surface electric field of the THz transient and the static bias with the emission times from (a) and (c) marked in grey.}
\end{figure*}

\section{Computation of the phase space density distribution}
\label{par: C}
The phase space density distribution is calculated with trajectories around a common temporal delay and for different initial energies. From these trajectories, we extract the distance from the tip $s$ and the electron momentum $p=m_ev$. The elements in phase space are weighted by the initial energy spectrum. To account for the finite duration of the photoemission pulse, the NIR-induced photoemission is averaging over a Gaussian-shaped emission window of $30$ fs. \\
The phase space density distribution of the electron pulses strongly depends on the relative emission phase within the THz cycle. To illustrate the variety of possible electron dynamics, we analyze the time evolution of the electron pulse in phase space for three exemplary emission times, as shown in Fig. 9. \\
The phase space evolution is traced via snapshots taken at equidistant temporal delays of $\Delta t = 50$ fs (cf. Fig. A4b,c) or $\Delta t = 100$ fs (cf. Fig. 9a). Figure 9b depicts the evolution of the phase space density distribution at the photocurrent onset ($t_0=0.2$ ps). Here, the acceleration by the THz electric field is strongly increasing in time, which results in a caustic partial inversion of the phase space density distribution. For comparison, Fig. 9a shows the acceleration of electron pulses by the static field in phase space. Finally, in Fig. 9c, the phase space density distribution at the energy minimum is depicted. \\
At a distance of $s=1\:\mu$m from the surface, a comparison of the pulse duration with and without THz-field shows an electron pulse duration of $24$ fs and $82$ fs (FWHM), respectively. This result highlights the potential of the THz near-field control for generating tailored ultrashort electron pulses.

\bibliography{citations}{}

\end{document}